\documentclass[lettersize,journal]{IEEEtran}
\usepackage{amsmath,amsfonts}
\usepackage{algorithmic}
\usepackage{algorithm}
\usepackage{array}
\usepackage[caption=false,font=normalsize,labelfont=sf,textfont=sf]{subfig}
\usepackage{textcomp}
\usepackage{stfloats}
\usepackage{url}
\usepackage{verbatim}
\usepackage{graphicx}
\usepackage{cite}
\usepackage{multirow}
\usepackage{booktabs}
\usepackage{xcolor}
\usepackage{tabularx}

\begin{document}
\title{Shaping a Smarter Electromagnetic Landscape: \\IAB, NCR, and RIS in 5G Standard and Future 6G}

\author{Chao-Kai~Wen,~\IEEEmembership{Fellow,~IEEE},~Lung-Sheng~Tsai,~Arman~Shojaeifard,~\IEEEmembership{Senior~Member,~IEEE}, Pei-Kai~Liao,~Kai-Kit~Wong,~\IEEEmembership{Fellow,~IEEE},~and~Chan-Byoung~Chae,~\IEEEmembership{Fellow,~IEEE}

\thanks{{C.-K.~Wen} is with the Institute of Communications Engineering, National Sun Yat-sen University, Kaohsiung 80424, Taiwan, Email: {\rm chaokai.wen@mail.nsysu.edu.tw}.}
\thanks{{L.-S.~Tsai} and {P.-K.~Liao} are with the MediaTek Inc., Hsinchu, Taiwan, Email: {\rm Longson.Tsai@mediatek.com, pk.liao@mediatek.com}.}
\thanks{{A.~Shojaeifard} is with InterDigital, London EC2A 3QR, U.K., Email: {\rm Arman.Shojaeifard@InterDigital.com}.}
\thanks{{K.-K.~Wong} is with Department of Electronic and Electrical Engineering, University College London, UK, Email: {\rm kai-kit.wong@ucl.ac.uk}. He is also affiliated with Yonsei Frontier Lab., Yonsei University, Korea.}
\thanks{{C.-B.~Chae} is with Yonsei University, Seoul 03722, Korea, Email: {\rm cbchae@yonsei.ac.kr}.}
}

\markboth{IEEE Communications Standards Magazine}%
{Shell \MakeLowercase{\textit{et al.}}: A Sample Article Using IEEEtran.cls for IEEE Journals}

%% Remember, if you use this you must call \IEEEpubidadjcol in the second
%% column for its text to clear the IEEEpubid mark.

\maketitle

\begin{abstract}
The main objective of 5G and beyond networks is to provide an optimal user experience in terms of throughput and reliability, irrespective of location and time. To achieve this, traditional fixed macro base station deployments are being replaced by more innovative and flexible solutions, such as wireless backhaul and relays. This article focuses on the evolution and standardization of these advancements, which are shaping the electromagnetic landscape. Specifically, we explore Integrated Access and Backhaul (IAB) nodes, which offer a cost-efficient and agile alternative to fiber backhaul. We also discuss Network-Controlled Repeaters (NCRs) and the emergence of Reconfigurable Intelligent Surfaces (RIS) actively adapting the wireless environment. The article provides an overview of the 5G features and ongoing developments in 3GPP Release 18 related to these intelligent EM entities, highlighting the expected evolution of future wireless networks in terms of architecture, operations, and control signals.
\end{abstract}

\begin{figure*}[b]
\centering
\includegraphics[width=6.00in]{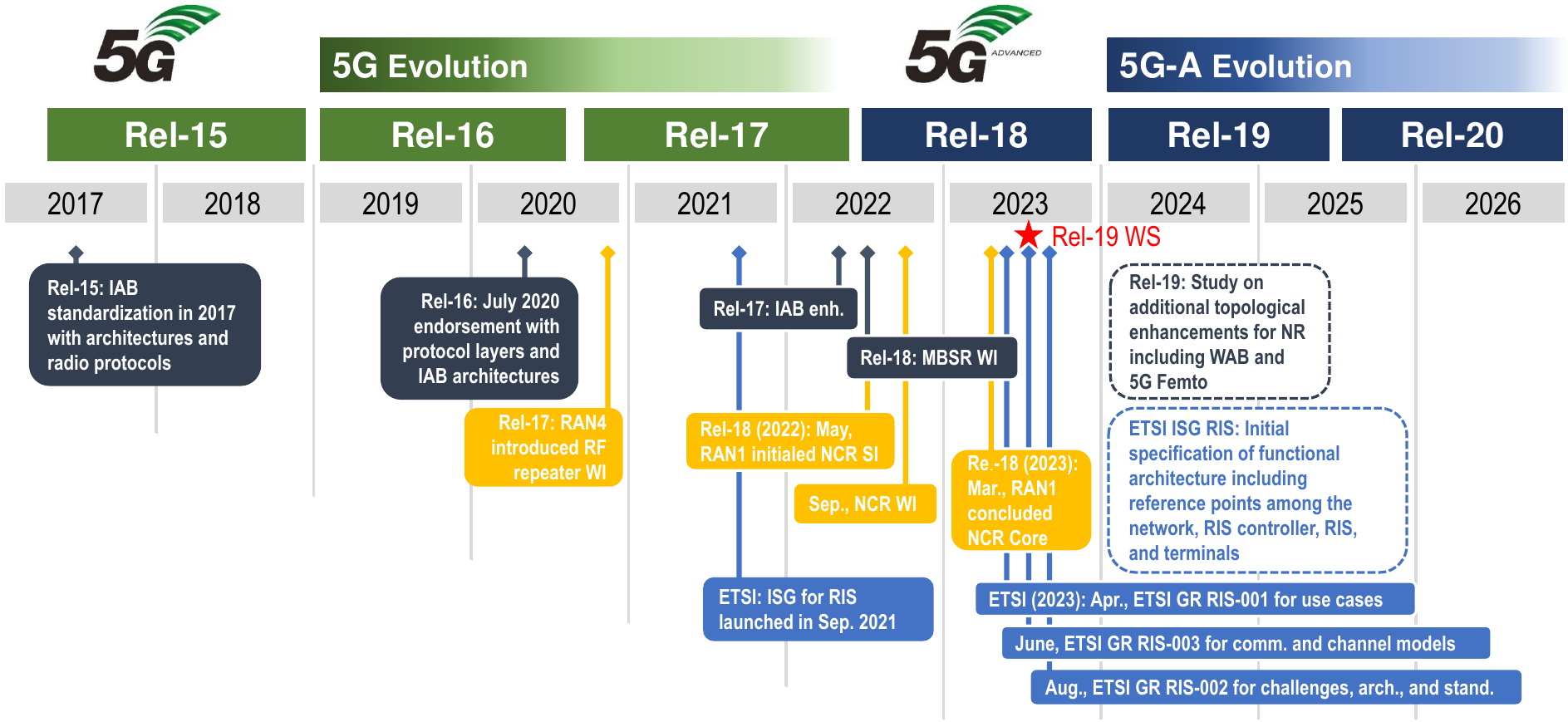}
\caption{5G standardization for IAB, NCR, and RIS.}
\label{fig:3GPP_Roadmap}
\end{figure*}

\section{Introduction}

\IEEEPARstart{T}{he} primary goal for 5G and beyond wireless networks is to satisfy user experience requirements for throughput and reliability, while simultaneously maintaining energy efficiency across all locations. Cellular networks traditionally relied on fixed macro base stations (BSs) for deployment. However, this approach may not offer sufficient coverage in densely populated urban areas or at higher frequencies. Network densification can address these coverage gaps, but creating new infrastructure from scratch is time-consuming and poses significant challenges due to cost, power sourcing, real estate permissions, regulatory approvals, and backhaul availability. In this context, wireless backhaul emerges as a feasible solution for enabling flexible and dense network deployments.

Wireless backhaul has been extensively studied, with the Long Term Evolution (LTE) relay playing a role in standardization efforts during LTE Release 10. However, due to the limited performance boost and the complexity of relay features, few operators adopted LTE relay. With the rise of dense 5G New Radio (NR) networks and advancements in beamforming technology, there is renewed interest in developing an Integrated Access and Backhaul (IAB) solution. This solution offers a faster and more cost-efficient alternative to fiber backhaul. Figure \ref{fig:3GPP_Roadmap} the 3GPP's roadmap for IAB development from 5G to 5G-Advanced. Studies on IAB architectures and radio protocols began with a Study Item (SI) within 3GPP Release 15 (Rel-15). Release 16 (Rel-16), detailed in Technical Specification TS 38.401 \cite{TS-38.401}, officially introduced a multi-hop NR-based IAB solution. Its objective was to use the existing 5G radio air interface for wireless backhaul while meeting specified electromagnetic (EM) compatibility requirements. Standardization efforts continued in Release 17 (Rel-17), with a focus on enhancing IAB network performance through topology adaptation, duplexing, and efficiency enhancements. Furthermore, Release 18 (Rel-18) introduced a Work Item (WI) to explore architectural and system-level enhancements for 5G networks with mobile BS relays mounted on vehicles, termed Vehicle Mounted Relay (VMR).

Apart from IAB, utilizing the decode-and-forward operation, employing amplify-and-forward RF repeaters is a simpler solution for improving network coverage. RF repeaters have been widely deployed in commercial 2G, 3G, and 4G networks. As 5G NR technology migrates to higher frequencies, signal propagation conditions may deteriorate, necessitating the critical implementation of RF repeaters. To accommodate this increasing demand, 3GPP RAN4 introduced specifications for RF repeaters in Rel-17 \cite{RP-202813}. These specifications lay out the RF and Electromagnetic Compatibility requirements for FR1 and FR2, ensuring compatibility in typical commercial environments.\footnote{In 3GPP, Frequency Range 1 (FR1) includes 410-7125 MHz, and Frequency Range 2 (FR2) includes 24.25-52.6 GHz.}

Although RF repeaters are cost-effective for expanding network coverage, their ability to support network performance-enhancing features such as adaptive spatial beamforming, dynamic gain and power adjustments, flexible downlink (DL)/uplink (UL) configurations, and ON-OFF stages is limited. To address these limitations, 3GPP initiated a SI on Network-Controlled Repeaters (NCRs) in Rel-18. NCRs inherit the amplify-and-forward operation of RF repeaters but also receive \emph{side control information} from the 5G gNB to function more efficiently. This SI was finalized in August 2022 with TR38.867 \cite{3GPP-38-867}, where RAN1 pinpointed the essential side control information necessary for NCR functionality. This was followed by the NCR WI \cite{RP-222673}, detailing side control information for beamforming, UL-DL Time Division Duplex (TDD) operations, and ON-OFF protocols. Figure~\ref{fig:3GPP_Roadmap} provides an overview of 3GPP's NCR roadmap.

Reconfigurable Intelligent Surfaces (RIS) are an emerging technology that leverages reconfigurable surface technology to adapt to the propagation environment. Primarily using passive components and eliminating the need for expensive active components like power amplifiers, RIS provides benefits such as lower hardware costs, reduced energy consumption, and versatile deployment options on various structures like walls, buildings, and lamp posts. Compared to IAB and NCR, RIS stands out due to its cost-effectiveness and adaptability.

As for standardization, RIS has not yet been studied in 3GPP. While some companies proposed including RIS as a SI in 3GPP for Rel-18, most considered it premature and suggested exploring it for 6G technology instead. Consequently, the proposal was not approved for Rel-18. To address standardization, the ETSI Industry Specification Group (ISG) for RIS was established in September 2021. It serves as the pre-standardizing group for RIS, aiming to define use cases, deployment scenarios, and requirements to establish global standardization. Their goal is to enable dynamic control over radio signals, effectively transforming the wireless environment into an adaptable service. The group's progress includes the release of its first Group Report (ETSI GR RIS-001 \cite{ETSI-GR-RIS-001}) in April 2023, which identifies pertinent use cases for RIS, followed by ETSI GR RIS-003 \cite{ETSI-GR-RIS-003} in June 2023, discussing communication and channel models, and ETSI GR RIS-002 \cite{ETSI-GR-RIS-002} in August 2023, which delves into technological challenges, architectural considerations, and their implications for standardization. Figure~\ref{fig:3GPP_Roadmap} showcases ETSI's progress on RIS.

With the standardization of IAB, NCR, and RIS, future networks are expected to incorporate a variety of network nodes to optimize performance, reduce costs, and minimize power consumption \cite{Flamini-TAP22}. 3GPP is currently advancing into the second phase of 5G standardization, known as 5G-Advanced, which builds upon the foundational 5G baseline established in 3GPP Releases 15, 16, and 17. The 3GPP Release 19 RAN workshop (Rel-19 WS), held on June 15-16, 2023, garnered significant global interest, with contributions from over 80 different companies and organizations. 3GPP Rel-19, which commenced in 2024, will study advanced network topology. In certain use cases, IAB, NCR, and RIS share similar deployment scenarios. Therefore, the progression of IAB and NCR within existing 5G and the ongoing 5G-Advanced developments provide mutual references. Additionally, their developments establish a foundational reference for the future standardization of RIS. This article aims to provide an overview of the 5G NR features pertinent to IAB and NCR, focusing on their anticipated evolution in architecture, operations, and control signals. Additionally, it offers insights into the current status and development of RIS as per the ETSI ISG RIS.

\section{IAB}

\begin{figure*}
\centering
\includegraphics[width=6.5in]{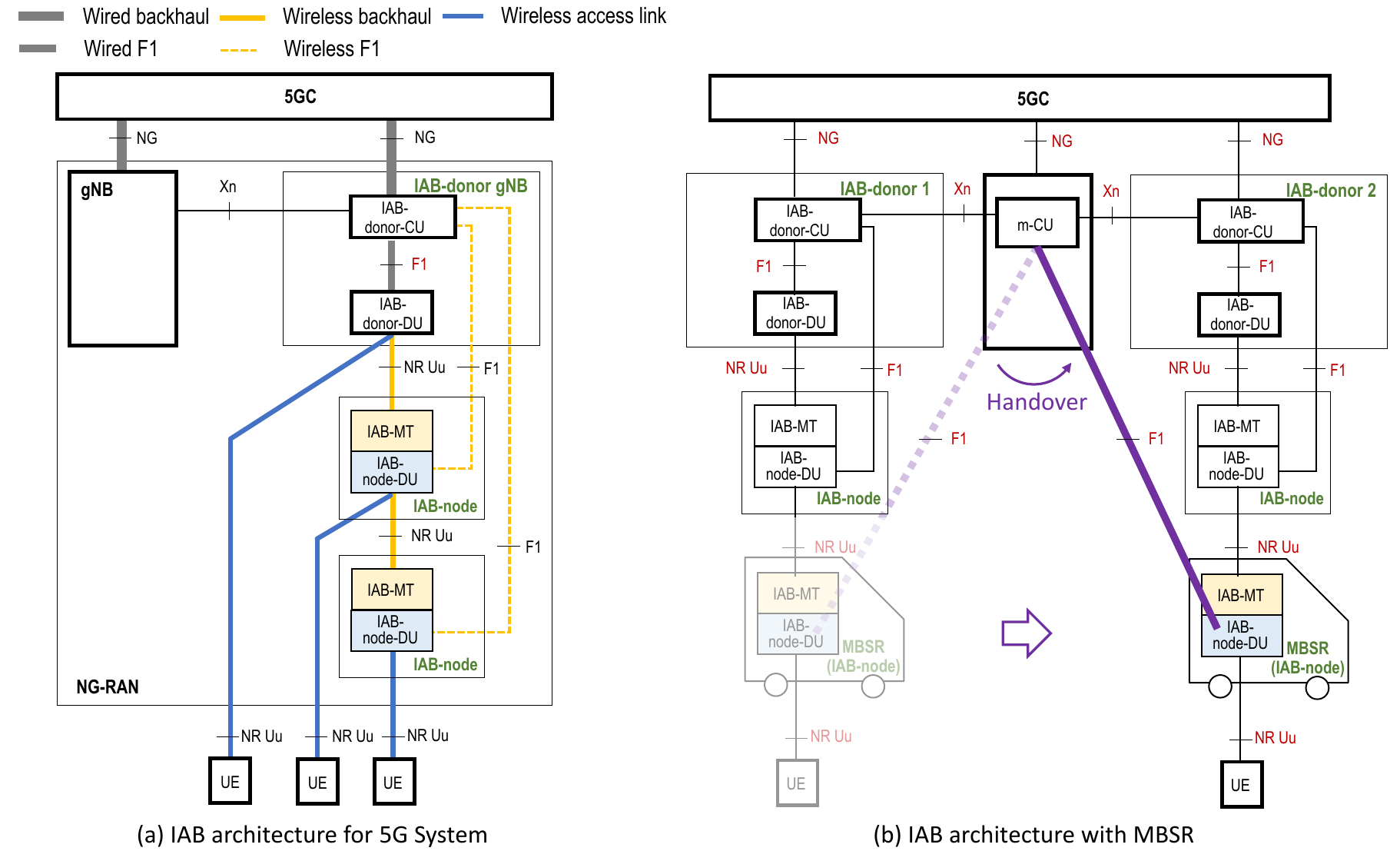}
\caption{IAB architectures for 5G system and MBSR.}
\label{fig:IAB_architecture}
\end{figure*}

Rel-16 introduced a multi-hop NR-based IAB solution, aimed at reusing the existing 5G radio air interface. This section explores the architecture of IAB, covering its network topology, resource allocation mechanisms, and recent enhancements, along with an outlook on future developments.

\subsection{IAB Architecture}

The architecture of IAB, depicted in Figure \ref{fig:IAB_architecture}(a), defines two types of network nodes \cite{TS-38.401}.

\subsubsection{IAB-donor} In the IAB network, the IAB-donor acts as the central control node, consisting of a centralized unit (IAB-donor-CU) and a distributed unit (IAB-donor-DU). These units are connected through a \emph{wired} F1 interface. The IAB-donor-DU handles the lower protocol layers (PHY, MAC, and RLC), while the IAB-donor-CU oversees the upper protocol layers (PDCP and SDAP/RRC).\footnote{Abbreviations: Physical (PHY), Medium Access Control (MAC), Radio Link Control (RLC), Packet Data Convergence Protocol (PDCP), and Service Data Adaptation Protocol (SDAP) / Radio Resource Control (RRC).} This division permits time-sensitive functions to be conducted by the IAB-donor-DU, which is closer to the served nodes, and the remaining functions by the IAB-donor-CU, which has superior processing capacity \cite{IAB-Overview22,IAB-chae}.

\subsubsection{IAB-node} The IAB-node is a \emph{wireless} relay node within the IAB network, tasked with initiating access to the parent node above it and providing access services to child nodes below it. As such, the IAB-node comprises two functional modules: the IAB Mobile Termination (IAB-MT) and the IAB-node-DU. The IAB-MT links an IAB-node with the DU of the parent/upstream node, which can either be an IAB-donor or another IAB-node. The IAB-node-DU (or IAB-donor-DU) caters to the user equipments (UEs) and potentially downstream IAB-nodes in instances of multi-hop wireless backhauling.

IAB-nodes function as Layer-2 regenerative relays, decoding and re-encoding each received packet prior to retransmission. This mechanism applies to packets received from the IAB-donor, UEs, or other IAB-nodes.

\subsection{IAB Network Topology}

Figure \ref{fig:IAB_architecture}(a) depicts the topology of an IAB network. The network originates from an IAB-donor, which serves as the central control node and traffic convergence point between the access and core network. The network expands in a tree-like structure through wireless links, connecting multiple intermediate nodes or IAB-nodes. These IAB-nodes provide UE access services and enable multi-hop transmissions, directing backhaul traffic to their respective child or upstream nodes.

The IAB-donor-CU establishes connections with all DU components in the IAB topology using wired or wireless F1 interfaces. It manages traffic to and from the core network while coordinating the operations of the entire IAB topology. For its child IAB-nodes, the IAB-donor provides wireless backhaul links within its radio coverage. From the perspective of UEs, there is no distinction between IAB-nodes, IAB-donors, and regular NR BSs. They all provide access links to UEs within their radio coverage through the IAB-DU module. The IAB-MT module, similar to a UE with a subset of UE functions, enables the IAB-node to connect to its parent DU node through the NR air interface.

During the initial power-up of an IAB-node, it needs to access the IAB network as a UE, acquire an IP address, and establish a wireless F1 interface between the IAB-node-DU and the IAB-donor-CU. Once the IAB-node-DU is configured, it can operate as a relay node within the network topology.

In a single-hop transmission scenario, the UE directly receives network services through the access link provided by the IAB-donor. However, in a multi-hop transmission scenario, the UE accesses the network via a neighboring IAB-node and utilizes the wireless backhaul function for multi-hop transmissions. The uplink transmission is relayed to the IAB-donor through multiple wireless backhaul links before reaching the core network, and the downlink transmission follows the reverse process.

\begin{table}[!t]
\caption{Comparisons of IAB and NCR.}\label{tab:IAB_vs_NCR}
\centering
\begin{tabular}{|l|c|c|}
\hline
 & \textbf{IAB} & \textbf{NCR} \\
\hline
3GPP Stage & Part of Rel-16 & WI concluded in Rel-18 \\
\hline
A\&B Links & Out-of-band/In-band & In-band \\
\hline
Architecture & Multi-hop & Single-hop \\
\hline
Deployment & Stationary/Mobile & Stationary \\
\hline
Operation & Decode-and-forward & Amplify-and-forward \\
\hline
Duplex Mode & HD/FD & FD \\
\hline
UE Transparency & Not transparent & Transparent \\
\hline
Power Saving & Specifics vary & Uses ON-OFF information \\
\hline
\end{tabular}
\end{table}

\subsection{IAB Resource Allocation Mechanism}

The IAB supports both out-of-band and in-band backhauling, with the latter using the same frequencies for NR backhaul and access links. In-band IAB may experience cross-link interference. To address this issue, various half-duplex (HD) multiplexing schemes have been developed. 3GPP Rel-16 primarily adopts Time-Division Multiplexing (TDM) for wireless resource allocation in IAB networks. In TDM, the IAB-node's parent link is allocated different time slots compared to its child links, thereby avoiding simultaneous transmissions and receptions by co-located MT and DU to prevent interference. 3GPP Rel-17 introduces other HD schemes like Frequency-Division Multiplexing (FDM) and Space-Division Multiplexing (SDM).

FDM assigns different frequencies to backhaul and access links, while SDM uses beamforming with multiple antennas for spatial separation. However, SDM does not always eliminate cross-link interference due to non-narrow beam widths and is often combined with TDM or FDM for optimal performance. Despite their utility, these HD schemes have limitations. TDM can cause relay delays, and FDM demands more spectrum. To improve spectral efficiency and reduce latency, 3GPP Rel-17 proposes a full-duplex (FD) mode. FD allows simultaneous transmission and reception on the same frequency in IAB-nodes. This, however, introduces significant self-interference, necessitating a strong self-interference cancellation mechanism for effective full-duplex operation in IAB networks \cite{IAB-chae}.

\subsection{Outlook}

Rel-16 ratified IAB, with enhancements continuing through Releases 17--18. Broadly, wireless backhaul enables the deployment of mobile cells, with Mobile Base Station Relays (MBSRs) often installed in vehicles like buses or trains, known as VMRs, to provide coverage for UE within or near the vehicle. 3GPP Rel-18 focuses on architecture enhancements for MBSRs \cite{TR-23.700}. MBSRs operate within a single-hop topology framework, where the lower IAB-node aligns with the MBSR. An intermediate IAB-node may exist, provided it does not function as an MBSR, as shown in Figure~\ref{fig:IAB_architecture}(b). Due to the mobility of an MBSR over a wide area, it may need to switch its IAB-donor. This can disrupt the upper protocol layer connections of the served UEs, even when the UEs are stationary inside the vehicle. To mitigate these disruptions, the introduction of a dedicated mobile control unit (m-CU) is proposed. The m-CU has an Xn connection to the donor gNB. By ensuring that the MBSR's DU is under the service of the m-CU, the MBSR can move across a larger RAN coverage area without requiring a change in the m-CU. This mobility allows the MBSR between IAB-donors to remain transparent to the connected UEs, as long as the control remains within the same m-CU. As illustrated in Figure \ref{fig:IAB_architecture}(b), during the MBSR's mobility between IAB-donors, the F1 interface between the MBSR and m-CU is preserved by transferring the F1 connection from the source to the target IAB-donors.

Rel-18 also addresses other pertinent challenges associated with MBSRs, such as implementing location services for UEs accessing the network through mobile or roaming MBSRs, ensuring accurate Cell ID/Tracking Area Code information despite MBSR movements, and developing efficient controls for managing UE access to the 5G network via MBSRs. Geographic constraints and legacy UE support are also factored in. The extension of IAB application scenarios to various domains is expected, including non-terrestrial network (NTN)-based backhauling, urban air mobility, public safety, and disaster recovery scenarios. These applications triggered a study of Wireless Access Backhaul (WAB) at Rel-19.

\section{NCR}

As another network node to improve coverage in 3GPP, the scope of NCRs, as detailed in the NCR WI \cite{RP-222673}, is more narrowly defined compared to IAB. According to TR 38.867 \cite{3GPP-38-867}, NCRs primarily focus on the following scenarios and assumptions:
\begin{itemize}
\item[-] NCRs are \emph{in-band} RF repeaters used to extend network coverage on FR1 and FR2 bands.
\item[-] Only \emph{single-hop stationary} NCRs are considered.
\item[-] The NCR is \emph{transparent} to the UE.
\item[-] The NCR can maintain the gNB-repeater link and the repeater-UE link \emph{simultaneously}.
\end{itemize}
Table \ref{tab:IAB_vs_NCR} provides a summary of the comparisons between IAB and NCR. In the subsequent subsections, we explore the architecture and functionalities of NCRs in more detail.

\subsection{NCR Architecture}

\begin{figure}
\centering
\includegraphics[width=3.5in]{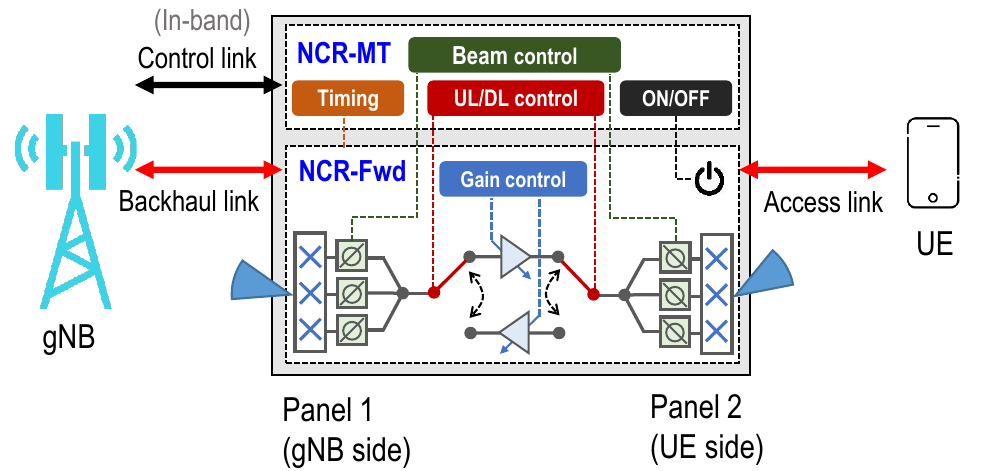}
\caption{NCR architecture.}
\label{fig:NCR}
\end{figure}

As shown in Figure \ref{fig:NCR}, the NCR consists of two functional entities \cite[Sec.~5]{3GPP-38-867}: the NCR-MT and the NCR-Forwarding (NCR-Fwd).
The NCR-MT, functioning similarly to the IAB-MT, connects to the gNB via a Control link (C-link) using the NR Uu interface. It supports a subset of UE functions to link with its parent gNB as a standard device while being identified as an NCR within the network. Furthermore, it manages side control information exchange to oversee the NCR-Fwd operations. Conversely, the NCR-Fwd serves purely as an RF repeater, relaying UL/DL RF signals between the gNB and UEs across the backhaul and access links. Its operation is guided by side control information relayed by the NCR-MT from the gNB. In summary, compared to the IAB-node, the NCR solely deciphers control information pertinent to itself and ignores all control, data, or signals meant for UEs.

The NCR-Fwd can be implemented with two sets of panel antennas (one for the backhaul link and the other for the access link) and one RF amplifier (see Figure \ref{fig:NCR}). It amplifies and beamforms the signal before forwarding it in the DL or UL direction. Beamforming techniques allow adjusting the reception and transmission directions. The NCR-Fwd module primarily focuses on signal amplification and (analog) beamforming, eliminating the need for advanced digital receiver or transmitter chains. The performance requirements for beamforming antennas in the NCR-Fwd are not as high as those for macro BS or IAB-node antennas. Cost-effectiveness and ease of manufacturing are prioritized.

The NCR-MT and NCR-Fwd can operate in the same or different frequency bands. However, at least one carrier used by the NCR-MT must operate within the frequency band being forwarded by the NCR-Fwd, which serves as the baseline.

\subsection{Side Control Information}
TR 38.867 \cite[Sec.~6]{3GPP-38-867} explores various forms of side control information, including but not limited to beam information, timing information, UL-DL TDD configuration, and ON-OFF information. We detail their signaling protocols in the subsequent subsections.

\subsubsection{Beam Information}

Deploying 5G NR in high-frequency bands necessitates beamforming capabilities in NCRs for optimal performance. The distinct characteristics of backhaul and access links allow for tailored beamforming mechanisms in the side control information:

\emph{Backhaul link and C-link}---Given the NCR's stationary nature, it can utilize fixed or adaptive beamforming on the backhaul and C-links to cope with varying conditions.
In the baseline scenario where the NCR-MT and NCR-Fwd operate within the same frequency band, it is expected that the C-link and backhaul link will experience similar large-scale channel characteristics. Therefore, the same transmission configuration indicator states used for the C-link can also be applied to the NCR-Fwd for beamforming in the backhaul link.
Furthermore, if adaptive beams are employed for both the C-link and backhaul link, the determination and indication of the backhaul link beams can be accomplished through new signaling provided by the gNB. In the absence of indication via the new signaling, the backhaul link beam can be determined based on a predefined rule.

\begin{figure}
\centering
\includegraphics[width=3.5in]{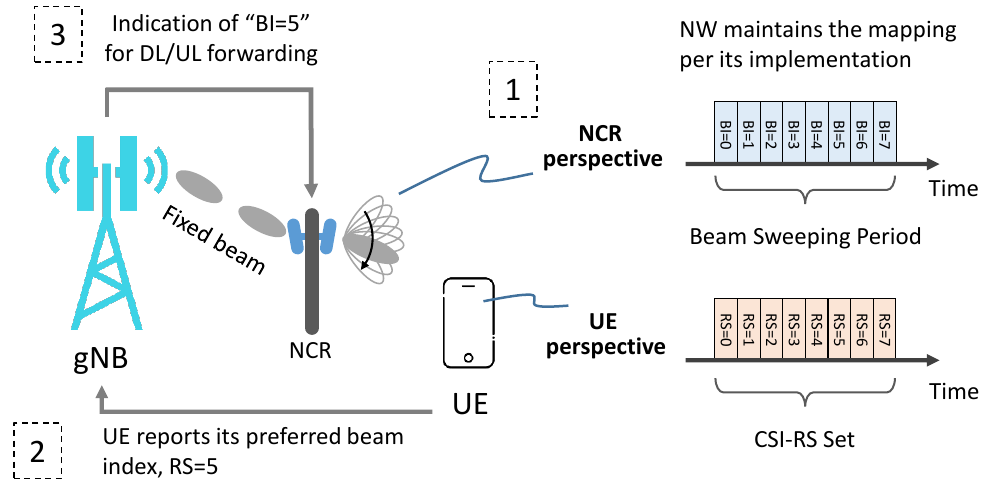}
\caption{Beam indication for access link.}
\label{fig:Beam_Indication}
\end{figure}

\emph{Access link}---In the context of the access link, dynamic beam steering towards users leads to improved SINR performance compared to fixed-beam solutions, particularly benefiting cell edge users \cite[Sec.~9]{3GPP-38-867}. Thus, providing beam information to guide the NCR's access link behavior is advisable.
The access link for NCR-Fwd is identified by a beam index, supporting both dynamic and semi-static indications. The dynamic indication allows for rapid adaptation to changing conditions, such as user mobility, whereas the semi-static indication holds a stable configuration with infrequent adjustments. Time domain resources must be explicitly linked to these beam indications. Specifically, the gNB specifies which time slots or symbols are to be allocated for a particular beam or set of beams to manage NCR-Fwd operations.
For the DL/UL of the access link in NCR-Fwd, beam correspondence is assumed. That is, the DL and UL beams on the access side that are paired with each other are assigned the same beam index.

To illustrate a beam indication mechanism for the access link, Figure \ref{fig:Beam_Indication} serves as an example. Here, the gNB guides the NCR through beam control information with Beam Index (BI) to perform periodic beam sweeping during the beam training phase. The gNB may not have full knowledge of the NCR's beam characteristics but knows the NCR's capability regarding the number of supported beams; meanwhile, the UE cannot differentiate between beams originating from the gNB or the NCR. Based solely on the received Reference Signals (RSs), specifically Channel State Information RS (CSI-RS), for beam quality measurement, the UE reports its preferred beam index, such as RS=5. From the UE's perspective, it cannot know which gNB's Tx beam and which NCR's Tx beam are applied for the transmission of each RS. In contrast, the gNB has full control over which gNB's Tx beam and which NCR's beam index are adopted for the transmission of RS=5. Subsequently, for subsequent data transmission, as depicted in Figure 4, the gNB indicates BI=5 as the NCR's Tx beamforming for forwarding.

\subsubsection{Timing Information}

The NCR's timing is assumed to follow these guidelines:
\begin{itemize}
\item[-] When internal delay is not considered: The NCR-Fwd's DL reception timing is synchronized with the NCR-MT's DL reception timing. Likewise, the NCR-Fwd's UL transmission timing is synchronized with the NCR-MT's UL transmission timing.

\item[-] When internal delay is considered: The NCR-Fwd's DL transmission timing occurs after an internal delay subsequent to the NCR-MT's DL reception timing. Conversely, the NCR-Fwd's UL reception timing precedes the NCR-MT's UL transmission timing by an internal delay.
\end{itemize}

\subsubsection{UL-DL TDD Configuration Information}
To avoid introducing significant signaling overhead in NCR, a semi-static TDD UL/DL configuration is supported for the C-link, backhaul link, and access link. To mitigate cross-link interference, the NCR-Fwd has its default behavior set to OFF on the flexible symbols/slots in the semi-static configuration. Additionally, for simplicity, the same TDD UL/DL configuration is assumed to be used for both the backhaul link and access link.

\subsubsection{ON-OFF Information}
Using ON-OFF information allows NCRs to be turned off when not needed, resulting in power savings. TR 38.867 \cite[Sec.~9]{3GPP-38-867} highlights that ON-OFF information not only saves power but also helps NCRs mitigate interference for high SINR users while maintaining performance for low SINR users, leading to an improved network experience. Therefore, ON-OFF information is recommended for NCRs to control the behavior of NCR-Fwd, providing benefits for network performance. By default, the NCR-Fwd is assumed to be OFF unless explicitly or implicitly indicated by the gNB.

\subsection{Outlook}

As Table \ref{tab:IAB_vs_NCR} illustrates, NCR demonstrates a narrower scope compared to IAB, primarily due to its early stage of application and specific design objectives. Further enhancements in the side control information for NCR are anticipated. For instance, at the Rel-19 WS, some companies highlighted the importance of power control in enhancing NCR performance. Additionally, strategies such as beamforming improvements, refined scheduling techniques, and the use of diverse frequencies for the backhaul link, may also be explored.

A significant trend observed in Rel-18 is the growing transition towards a more UE-centric architecture within the evolution of massive MIMO. One such example is the coherent joint transmission for multi-transmission/reception point. This shift is motivated by the increasing quantity of UEs, which induces complexities in network management. Consistent with this trend, the concept of a UE-controlled repeater (UCR) has been proposed at the Rel-19 WS \cite{RWS-230111}. The UCR employs an amplify-and-forward Layer-1 forwarding with frequency translation mechanism, which relays received signals to the UE using a different frequency band. This method enables end-user-centric collaborative MIMO, as described in \cite{Tsai-TAP22}. Here, multiple fixed or portable devices cooperate to create a rich array of antennas, thereby resulting in substantial performance enhancements.

\section{RIS}

RIS has not yet been designated as a SI by 3GPP. According to ETSI GR RIS-001 \cite{ETSI-GR-RIS-001}, RIS is defined as:

\vspace{0.2cm}
\noindent \emph{``RIS is a new type of system node with reconfigurable
surface technology, which can adapt its response according to
the status of the propagation environment through
control signaling.''}

\vspace{0.2cm}
\noindent In particular, RIS manipulates incoming wireless signals through techniques like reflection, refraction, absorption, and backscattering. It includes active, passive, and hybrid designs.\footnote{Active RIS uses powered RF circuits, passive RIS employs cost-effective elements to modify EM fields, and hybrid RIS merges reflection with signal sensing for communication enhancement while retaining passive RIS's energy efficiency.}

\begin{figure}
\centering
\includegraphics[width=3.5in]{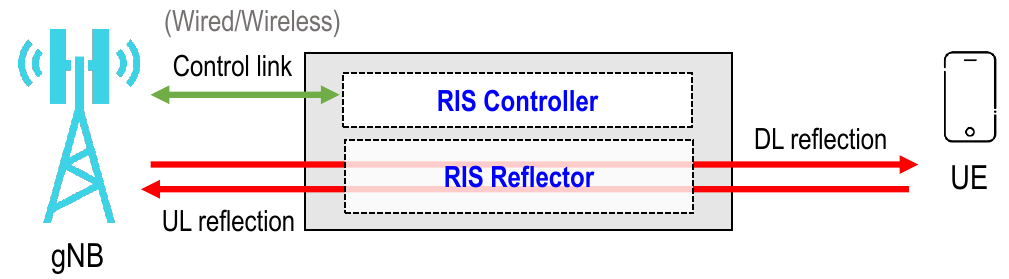}
\caption{Potential RIS architecture \cite{Yuan-COMMag22}.}
\label{fig:RIS}
\end{figure}

\subsection{RIS Architecture}

RIS can be modelled as a combination of a RIS controller and a RIS panel, as detailed in ETSI GR RIS-002 \cite[Fig. 5.1-1]{ETSI-GR-RIS-002}. The panel is equipped with elements capable of altering the characteristics of incoming radio waves, either reflecting or redirecting them based on the panel's design. The RIS controller not only adjusts these elements to manipulate the waves but also processes control signals from other network nodes. Functionally, RIS could resemble NCR, leading to suggestions to adapt the existing Rel-18 NCR architecture for RIS applications. For example, the adoption of an NCR-like architecture with separate RFs for control and reflection to increase design flexibility has been proposed \cite{Yuan-COMMag22}, as depicted in Figure~\ref{fig:RIS}. Unlike NCR and IAB, RIS may not include a MT unit. Instead, it features a streamlined control unit, connected via a wired or wireless control link, and uses a single reflective panel instead of separate receive and transmit antenna panels.

While the NCR-like architecture for RIS shows promise, it is also important to consider alternative architectures due to RIS's diverse deployment scenarios. The ETSI GR RIS-001 \cite{ETSI-GR-RIS-001} highlights key environments for RIS deployment, including indoor, outdoor, and hybrid settings. RIS can be deployed in either fixed or nomadic manners. In a fixed deployment, the RIS is attached to a static structure, such as a building wall, creating a largely static radio channel with a stationary BS. On the other hand, the nomadic deployment model permits mounting the RIS on moving platforms like trains or vehicles, allowing dynamic changes in its location or orientation. Therefore, network operators and service providers need to devise specific control strategies for each type of deployment, considering that NCR-like approaches might not always be applicable. As expounded in \cite{ETSI-GR-RIS-002}, RIS control methodologies include:
\begin{itemize}
\item \textbf{Network-controlled RIS}: The network directly provides configuration commands based on measurements from UEs and/or RIS, with the RIS controller under network jurisdiction.

\item \textbf{Network-assisted RIS}: The RIS controller, which could be part of the network or a third party, uses UE and/or RIS measurements with network input to adjust the RIS settings.

\item \textbf{Standalone RIS}: This controller independently dictates RIS configurations based on UE and/or RIS measurements.

\item \textbf{UE-controlled RIS}: Configuration control lies with the UE, which instructs the RIS through its controller.

\item \textbf{Hybrid-controlled RIS}: The RIS controller is bifurcated into remote and local segments. The remote segment is part of the network or a third party, while the local is embedded in the RIS controller.
\end{itemize}

\subsection{RIS Requirements}

RIS distinguishes itself from IAB and NCR in terms of cost efficiency and energy conservation.
Specifically, according to ETSI GR RIS-001 \cite{ETSI-GR-RIS-001}, RIS requirements include:

\begin{itemize}
\item {\bf Hardware Cost}: RIS should be more cost-effective than IAB and NCR, factoring in production and component costs.

\item {\bf Ease of Deployment and Maintenance}: Its deployment should be straightforward, regardless of whether fixed or wireless backhaul is used. Maintenance including fault detection and software updates should be easy.

\item {\bf Signal Power Boosting}: The RIS's ability to enhance signal power is influenced by its size, number of elements, and configurability.

\item {\bf Reconfigurability}: RIS should rapidly adapt to changes, with attention to the quantity of elements that can be reconfigured simultaneously.

\item {\bf Interoperability and Regulatory Compliance}: It must integrate seamlessly with existing networks and adhere to EM exposure regulations.

\end{itemize}

By meeting these requirements, RIS facilitates integration into various applications, extending beyond the coverage-focused roles of IAB and NCR. Its uses range from improving coverage to enabling wireless power transfer, supporting ambient backscatter communications, enhancing positioning accuracy, and strengthening secure communication. This contributes to the development of a ubiquitous intelligent network.

\subsection{Outlook}

The ETSI ISG RIS completed its first phase, spanning two years, concentrating on exploring technological potential, validation, and standardization requirements. This phase concluded with the publication of three GRs. The organization will now enter the second phase, targeting the initial specification of the functional architecture.
During the Rel-19 WS, various companies and institutes proposed a study or feasibility phase for RIS, which can also reflect a guide for this specification.
The shared perspectives include:

\begin{itemize}
\item \textbf{Channel Modeling}: Unlike IAB and NCR, where the channel models between two nodes can be modeled like a transmit-receive pair using the TR 38.901 \cite{TR-38.901} channel model, RIS is seen as a reconfigurable cluster, leading to a more complex channel model. Industry proposals highlight the need to examine channel modeling for RIS, taking into account a range of propagation effects such as reflection, refraction, absorption, and scattering. Models should represent diverse environments---indoor, outdoor, outdoor-to-indoor, and line-of-sight (LoS)/Non-LoS situations---as well as fluctuations at both large and small scales, radiation characteristics, and behaviors in both near and far fields. ETSI GR RIS-003 \cite{ETSI-GR-RIS-003} highlights the necessity of creating models that strike a balance between detailed complexity and practical accuracy.

\item \textbf{Use Cases and Deployment Scenarios}: The companies generally proposed studying various use cases, deployment scenarios, and operation modes of RIS, with an emphasis on their potential for improving network coverage and communication performance. This includes investigating RIS's role in both FR1 and FR2 frequency bands and exploring the potential for cooperative transmission and interference mitigation.

\item \textbf{RIS Architecture and Control Signaling}: The companies proposed advanced control information and signaling for RIS, including power control, signal operation, and beam management. This highlights the need to study how RIS integrates with and affects existing technological frameworks. Similar points are also discussed in ETSI GR RIS-002 \cite[Sec. 7]{ETSI-GR-RIS-002}.
\end{itemize}

\section{Conclusion}
This article provided a comprehensive overview of the 3GPP standardization efforts for IAB, NCR, and RIS. IAB enhances coverage and capacity using Layer-2 capable decode-and-forward relays, while NCR extends coverage through simpler amplify-and-forward repeaters. RIS, on the other hand, manipulates signals using cost-effective, energy-efficient reconfigurable unit-cells. As these network nodes standardize and evolve within 3GPP, as depicted in Figure \ref{fig:3GPP_Roadmap}, future wireless networks are expected to incorporate a mix of gNBs and IABs, creating a network of macro and small cells. NCR and RIS will improve coverage and UE connectivity, leading to a sophisticated wireless communication landscape.


\end{document}